\documentclass[showpacs,superscriptaddress]{revtex4}
\usepackage{graphicx,amsmath,amssymb,amsfonts,bbm,latexsym,color,dcolumn,epsf}

\newcommand{\be}{\begin{equation}}
\newcommand{\bea}{\begin{eqnarray}}
\newcommand{\ee}{\end{equation}}
\newcommand{\eea}{\end{eqnarray}}
\newcommand{\eq}[1]{Eq.~(\ref{#1})}

\begin{document}

\title{Vacuum polarization around stars: nonlocal approximation}
\author{Alejandro Satz}
\affiliation{School of Mathematical Sciences, University of Nottingham}

\author{Francisco D. Mazzitelli}
\affiliation{Departamento de F\'\i sica {\it Juan Jos\'e
Giambiagi}, FCEyN UBA, Facultad de Ciencias Exactas y Naturales,
Ciudad Universitaria, Pabell\' on I, 1428 Buenos Aires, Argentina}

\author{Ezequiel Alvarez}
\affiliation{Departament de Fisica Teorica. IFIC,CSIC -
Universitat de Valencia - Dr Moliner 50, E-46100 Burjassot
(Valencia) Spain}

\begin{abstract}
We compute the vacuum polarization associated with quantum
massless fields around stars with spherical symmetry. The nonlocal
contribution to the vacuum polarization is dominant in the
weak field limit,  and  induces quantum corrections to the
exterior metric that depend on the inner structure of the star.
It also violates the null energy conditions. We argue that
similar results also hold in the low energy limit of quantum
gravity. Previous calculations of the vacuum polarization in
spherically symmetric spacetimes, based on local approximations,
are not adequate for newtonian stars.
\end{abstract}

\maketitle

\section{Introduction}
In quantum field theory in curved spaces, the mean value of the
energy momentum tensor $\langle T_{\mu\nu}(x)\rangle$ plays a
main role. When evaluated around black holes or collapsing stars,
it contains the information about Hawking radiation.  As a source
in the semiclassical Einstein equations, it is responsible for
the quantum corrections to the metric.

It is well known that $\langle T_{\mu\nu}(x)\rangle$ is a nonlocal
 object, in the sense that its value at a given point $x$
may depend of the values of the metric in the past null cone of
$x$ (it may also depend on the topology of the manifold). This is
clear for example in cosmological situations, where the number of
particles created depend on the whole history of the scale
factor. It is less obvious for collapsing stars and black holes
(Hawking radiation only depends on the surface gravity of the black hole),
but in principle the exact $\langle T_{\mu\nu}(x)\rangle$ should
depend nonlocally on the metric.

Given a spacetime metric, it is in general difficult to compute
$\langle T_{\mu\nu}(x)\rangle$ exactly. It is even more difficult
to solve self-consistently the semiclassical Einstein equation
with $\langle T_{\mu\nu}(x)\rangle$ as a source, since this would
involve the knowledge of the vacuum polarization for at least a
family  of metrics. For these reasons, it is of interest to
develop approximation and numerical schemes. If the quantum field
is very massive, the DeWitt-Schwinger approximation is adequate
to describe the vacuum polarization \cite{sdw}. However, if the
mass is small or vanishes, the situation is more complex. There
are many papers dealing with this problem, starting from those
concerned with conformal fields in Schwarzschild spacetime
\cite{candelas,page,frolov,brown}. Analytic approximations have
been proposed for Schwarzschild, Reissner Nordstrom, and general
static spherically symmetric spacetimes, for both conformal and
non conformal fields \cite{ahs}. The approach of \cite{ahs}
contains as particular cases  the previous analytic
approximations \cite{page,frolov,brown} to $\langle
T_{\mu\nu}(x)\rangle $ for black holes. It has been shown
\cite{ahs} that the analytic approximation reproduces the exact
result for Schwarzschild spacetime with high accuracy in the case
of massless scalar fields (however this is not the case for
spinor fields, see \cite{a03}).

The different approximations to $\langle T_{\mu\nu}(x)\rangle$
depend of course on the quantum state of the field. The states
usually considered are the Hartle-Hawking, Unruh and Boulware
vacua \cite{vac}. The Hartle-Hawking  state describes a black hole
in equilibrium with its thermal radiation. The Unruh state is the
one that best mimics the gravitational collapse of a star. The
Boulware state, being  ill defined on the horizon, is unphysical
for black holes. However, it is  widely believed to describe the
quantum vacuum around a non-collapsed object like a star or a
planet. For these reasons, analytic approximations in the Boulware
state have been used to compute vacuum polarization around stars
\cite{his} and to check the validity of the energy conditions
\cite{visser1}.

In the approximations mentioned above, the energy momentum tensor
is determined by local functions of the metric and its
derivatives. However, as we already pointed out, on general
grounds one expects   $\langle T_{\mu\nu}(x)\rangle$ to be a
nonlocal function. In particular, in the weak field approximation,
a covariant perturbative approximation to $\langle
T_{\mu\nu}(x)\rangle$ is explicitly nonlocal \cite{horo, barvi}.
In this paper we will compute $\langle T_{\mu\nu}(x)\rangle $
around static spherical stars using this nonlocal approximation.
We will consider massless fields with arbitrary coupling to the
curvature in the Boulware quantum state. We will compare the
result with the analytic approximations, and show that the
nonlocal part dominates. Moreover, we will show that from the
nonlocal result one can derive the well known $1/r^3$- quantum
correction to the Newtonian potential \cite{newton,dono}.
Conversely, we will see that the nonlocal approximation can be
easily understood from this quantum correction and the
superposition principle, valid in the weak field approximation.

The paper is organized as follows. In the next section, as a
warm-up, we discuss the local and nonlocal approximations to
$\langle\phi^2 \rangle$. We expand the nonlocal approximation in a
multipolar expansion, which shows explicitly its dependence with
the internal structure of the star. We also show that there are
surface divergences on the boundary of the star unless the Ricci
tensor is sufficiently smooth. In Section III we compute $\langle
T_{\mu\nu}(x)\rangle$ in the weak field approximation for an
arbitrary, static, spherically symmetric star. We point out again
the dependence on the inner structure of the star and the
existence of surface divergences. Section IV contains some
applications: first we show that $\langle T_{\mu\nu}(x)\rangle$
violates both the null energy and the average null energy
conditions. Then we compute quantum corrections to the exterior
metric, and find that they depend on the internal structure of the
star. Section V contains a short discussion and conclusions.

Throughout this paper we use units in which $\hbar=c=1$, while retaining $G=l_p^2\neq 1$.
Our sign conventions are $---$ in the nomenclature of \cite{MTW}

\section{Nonlocal approximation for $\left\langle \phi^2\right\rangle$}

In this section we compute the quantity $\langle \phi^2\rangle$,
for a massless scalar field with arbitrary coupling ($\xi$) to the
curvature, in a weak background gravitational field.  We assume
the classical source that generates the field to be static and
non-relativistic, i.e., its stress energy tensor takes the form
$T_{\mu\nu}(x)=\rho(\textbf{x})\delta^0_{\mu}\delta^0_{\nu}$.

An expression for $\langle \phi^2\rangle$ can be obtained by
taking the coincidence limit of the Feynman propagator,
\begin{equation}
\left\langle \phi^{2}(x)\right\rangle=-\mathrm{Im}\,\Big(\lim_{x'\to x}G_{F}(x,x')\Big) ,
\end{equation}
and then renormalizing it.  With this purpose we first find the
Feynman Green function by solving the equation
\begin{equation}
\label{eq;KGfeynman} \left[\square_{x} + \xi
R(x)\right]G_{F}(x,x')=-\frac{1}{\sqrt{g(x)}}\,\delta(x-x') .
\end{equation}
Since we are working in a weak gravitational field, we assume that
both the metric and the Green function differ only slightly from
their Minkowski space counterparts:
$g_{\mu\nu}=\eta_{\mu\nu}+h_{\mu\nu}$ and
$G_F(x,x')=G_F^{(0)}(x,x')+G_F^{(1)}(x,x')$. Expanding
\eq{eq;KGfeynman} in these small quantities and transforming
Fourier in $x$, it is straightforward to solve it to first order
as \bea \label{eq;integralGF} G_F (x,x') &=& G_F^{(0)}(x,x') +
\frac{1}{(2\pi)^8} \int \mathrm{d}^4k  \int \mathrm{d}^4k' \int
\mathrm{d}^4\tilde{x} \frac{e^{ik(x-\tilde{x})}}{k^2}\Big[\xi
R(\tilde{x})-\bar{h}^{00}\,\partial_{\tilde{t}}^2\Big]
\frac{e^{ik'(\tilde{x}-x')}}{k^{\prime
2}} , \eea where $\bar h_{\mu\nu} = h_{\mu\nu} - \frac{1}{2}
\eta_{\mu\nu} \mbox{Tr} (h)$, and now $R(x)$ is the Ricci scalar
calculated to first order in $h_{\mu\nu}$.  After changing
variables to $p = k - k'$ and $q = k + k'$, the $q$-integrals in
\eq{eq;integralGF} can be performed by standard dimensional
regularization techniques.  Integrating in $d$-dimensional
$q$-space and taking the coincidence limit leads to
\begin{eqnarray}\label{phi2form}
 G_F^{(1)}(x,x)&=&\left(\xi-\frac{1}{6}\right)\Big[\frac{i}{8\pi^2}\frac{R(x)}{4-d}-\frac{i}
 {256\pi^6}\int\mathrm{d}^4p \int \mathrm{d}^4\tilde{x} \,\,R(\tilde{x})\, e^{ip(x-\tilde{x})}
 \ln \frac{p^2}{\mu^2}\Big] \nonumber\\
 &\equiv &\left(\xi-\frac{1}{6}\right)\Big[\frac{i}{8\pi^2}\frac{R(x)}{4-d}-\frac{i}
 {256\pi^6}\ln\frac{\square}{\mu^2}\, R\Big]\,\, ,
\label{int-square}
\end{eqnarray}
where $\mu$ is an introduced arbitrary mass scale.  The first term
in the square bracket in \eq{phi2form} is the divergent part of
the Hadamard propagator which must be removed from the result to
obtain a finite --and renormalized-- value for $G_F (x, x)$. The
second term, on the other hand, contains two pieces: one local and
arbitrary proportional to $R(x) \ln \mu^2$ and with no physical
significance, and other non-local and with physical meaning to
$\langle \phi^2 \rangle$.

Once dropped the $\frac{1}{d-4}$ divergent part in \eq{phi2form}, the integration in $\tilde t$ and $p^0$ for a
static situation yields
\begin{equation}
\label{eq;fourier} \left\langle
\phi^2(x)\right\rangle=\frac{1}{128\pi^5}\left(\xi-\frac{1}{6}\right)\int
\mathrm{d}^3p\,\,\tilde{R}(p)
\,e^{-i\,\textbf{p}\cdot\textbf{x}}\ln\frac{\left|\textbf{p}\right|^2}{\mu^2}
\equiv
\frac{1}{128\pi^5}\left(\xi-\frac{1}{6}\right)\ln\frac{-\nabla^2}{\mu^2}\,
R ,
\end{equation}
where $\tilde R (p)$ is the Fourier transform of $R(x)$.   This
result is finite everywhere for any smooth and asymptotically flat
$R(x)$.  Notice, however, that for the external region of the
source, i.e.~$R(x)=0$, \eq{eq;fourier} can be taken to a more
friendly form if the  d$^3 p$ integral is performed first in
\eq{phi2form}, namely
\begin{equation}
\label{eq;GFnoloc} \langle \phi^2 (x) \rangle
=-\frac{1}{32\pi^3}\left(\xi-\frac{1}{6}\right)\int d^3
\tilde{x}\,\frac{
R(\tilde{x})}{\left|\textbf{x}-\tilde{\textbf{x}}\right|^3} \ \ \
\ \ \ \ \ \ \ \ \mbox{(external region)}.
\end{equation}

As an application of the results in
Eqs.~(\ref{eq;fourier},\ref{eq;GFnoloc})  we now compute the
vacuum polarization for a spherical star of radius $R_0$ and
constant density $\rho_0$ (we assume $\rho_0 G R_0^2 \ll 1$ to
endorse the weak field approximation).  In this case the scalar
curvature takes the form $R(x) = 8\pi G \Theta (R_0 - r)$ and
\eq{eq;fourier} yields
\begin{equation} \label{eq;rhocte}
\left\langle\phi^2(r)\right\rangle=-\frac{G\rho_0}{2\pi}\left(\xi-\frac{1}{6}\right)\left
\{ \begin{array}{ll}
\ln[C(R_0^2-r^2)] & \quad r<R_0 \\
-\frac{2R_0}{r}+\ln\Big(\frac{r+R_0}{r-R_0}\Big) & \quad r>R_0 ,\\
\end{array} \right.
\end{equation}
where  $C$ is an arbitrary constant proportional to $\mu^2$.
Notice that this expression has a logarithmic divergence at the
surface of the star due to the discontinuity of the density; this
is because the Fourier transform of the Heaviside function falls
only as $p^{-1}$ when $p \to \infty$.  In effect, any star model
with a non-continuous $\rho$  at the surface will have a
logarithmic divergence in $\langle \phi^2  \rangle$ at the point
of discontinuity. On the other hand, for a continuous density the
Fourier transforms falls faster to zero and the divergence
disappears.  The surface divergence is similar to the divergences
that appear when computing the Casimir effect for perfect
conductors of arbitrary shape. See for instance \cite{deutch}.

It should be noted that the expression for $\left\langle
\phi^2\right\rangle$ outside the star in \eq{eq;rhocte}, depends
not only on the mass of the star but also on its radius; in fact,
expanding this expression for $r > R_0$ yields
\begin{equation}
\left\langle\phi^{2}(r)\right\rangle=-\frac{(\xi-\frac{1}{6})}{4\pi^2}\,\,\frac{MG}{r^3}
\left(1+\frac{3}{5}\frac{R_0^2}{r^2}+\cdots\right)\quad(r>R_0) ,
\end{equation}
where $M$ is the total mass of the star. This kind of dependence is general for any star model, as can be seen by replacing \eq{eq;GFnoloc} by its multipolar expansion,
\begin{equation}\label{eq;polenserie}
\left\langle \phi^{2}(r)\right\rangle= -\frac{G}{4\pi^2}\left(\xi-\frac{1}{6}\right) \sum_{n=1}^\infty \frac{M_n}{r^{2n+1}} ,
\end{equation}
which is exact in the external region for any spherically symmetric distribution of mass. The multipolar coefficients in \eq{eq;polenserie} are defined as
\begin{equation}\label{eq;defMn}
M_n=4\pi \int_0^{R_0}d\tilde{r}\,\rho(\tilde{r})\,\tilde{r}^{2n} ,
\end{equation}
so that $M_1=M$ and in general $M_n\sim M R_0^{2(n-1)}$. Therefore, the vacuum polarization in the external region depends at the sub-dominant order on the {\it internal structure} of the star.

In contrast to this internal structure dependence, we point out that the local approximation for $\left\langle \phi^2\right\rangle$ in static spherically symmetrical spaces previously developed \cite{mranderson}, for a massless field in the Schwarzschild metric leads to
\begin{equation}
\label{sandia}
\left\langle \phi^{2}(r)\right\rangle_{\mathrm{loc}}=\frac{M^2G^2}{48\pi^2r^3(2MG-r)} ,
\end{equation}
which, in the weak field approximation, goes as $M^2G^2/r^4$ since we are assuming $r \gg 2MG$ in the external region. A comparison between \eq{sandia} and \eq{eq;polenserie} for large $r$ gives
\begin{equation}\label{eq;complnl}
\frac{\left\langle \phi^{2}(r\to \infty)\right\rangle_{\mathrm{nonloc}}}{\left\langle \phi^{2}(r)\right\rangle_{\mathrm{loc}}}\sim \left(\xi-\frac{1}{6}\right)\frac{r}{MG} ,
\end{equation}
and hence, except in the conformal case $\xi = 1/6$, the nonlocal approximation is dominant.   Moreover, since the nonlocal expression in \eq{eq;rhocte} increases when $r \to R_0$ outside the star, the nonlocal expression is dominant everywhere in the external region.

We have therefore shown through an easy object, as $\langle \phi^2 \rangle$, how the quantum effects on a weak gravitational field can induce dominant nonlocal behaviors, providing a conceptual difference with previous works in the subject. Our next step is to use the light thrown by this calculation to obtain and study analogous results for a more complicated and important object, the stress-energy tensor.

\section{Nonlocal approximation for $\left\langle T_{\mu\nu}\right\rangle$}

A nonlocal formal expression for $\left\langle
T_{\mu\nu}\right\rangle$ in the weak field limit, similar in
character to our expression for $\left\langle \phi^2\right\rangle$
in Eqs. (\ref{int-square})-(\ref{eq;fourier}), has been found in
\cite{barvi,horo}. In the massless case it reads
\begin{equation}\label{Tmunuform}
 \left\langle T_{\mu\nu}(x)\right\rangle =-\frac{1}{32\pi^2}\Bigg\{ \frac{1}{2}
 \left[\left(\xi-\frac{1}{6}\right)^2-\frac{1}{90}\right]\times \ln\frac{\square}{\mu^2}
 H_{\mu\nu}^{(1)}(x) +\frac{1}{60}\ln\frac{\square}{\mu^2} H_{\mu\nu}^{(2)}(x)
 \Bigg\}\,\,\, ,
\end{equation}
where the tensors $H_{\mu\nu}^{(i)}$ are the two independent
higher-order tensors that appear in the Einstein equations when
derived from an action including $R^2$ and $R_{\mu\nu}R^{\mu\nu}$.
In the weak field limit they reduce to
\begin{eqnarray}\label{defHmunu}
H_{\mu\nu}^{(1)}&=& 4\nabla_{\mu}\nabla_{\nu}R-4g_{\mu\nu}\square R+O\,(R^2)\nonumber\\
H_{\mu\nu}^{(2)}&=&2\nabla_{\mu}\nabla_{\nu}R-g_{\mu\nu}\square
R-2\square R_{\mu\nu}+O\,(R^2)\,\, .
\end{eqnarray}

The action of the nonlocal operator $F(\square)=
\ln\frac{\square}{\mu^2}$ in Eq. (\ref{Tmunuform}) has been
described in detail in several papers (see for example
\cite{horo}). For time dependent situations it involves an
integral in the past null cone of $x$, and  therefore
$\left\langle T_{\mu\nu}\right\rangle$ is nonlocal and causal. On
the other hand, for time independent situations, it can be shown
that $F(\square)=F(-\nabla^2)$ \cite{dm2}.

From Eq. (\ref{Tmunuform}), assuming a static situation and
performing the relevant integrations, it is possible to derive a
nonlocal expression analogous to (\ref{eq;GFnoloc}), valid in the
external region where the classical source vanishes. It is given
by
\begin{equation}\label{eq;Tmununoloc}
\left\langle
T_{\mu\nu}(x)\right\rangle=\frac{1}{128\pi^3}\Bigg\{\left[\left(\xi-\frac{1}{6}\right)^2-
\frac{1}{90}\right]\int
d^3\tilde{x}\,\,\frac{H_{\mu\nu}^{(1)}(\tilde{x})}{\left|\textbf{x}-\tilde{\textbf{x}}
\right|^3} +\frac{1}{30}\int
d^3\tilde{x}\,\,\frac{H_{\mu\nu}^{(2)}
(\tilde{x})}{\left|\textbf{x}-\tilde{\textbf{x}}\right|^3}\Bigg\}\,\,
.
\end{equation}
Inserting the definitions Eq.(\ref{defHmunu}) into this equation,
after integration by parts we obtain
\begin{eqnarray} \label{Tmunuderivado}
\left\langle T_{00}\right\rangle &=& \frac{1}{32\pi^3}\left[\left(\xi-\frac{1}{6}\right)^2-
\frac{1}{90}\right]\nabla_x^2\int d^3\tilde{x}\,\,\frac{R(\tilde{x})}{\left|\vec{x}-
\vec{\tilde{x}}\right|^3}\nonumber\\
\left\langle T_{0i}\right\rangle &=& 0 \nonumber\\
\left\langle T_{ij}\right\rangle &=& \frac{1}{32\pi^3}
\left[\left(\xi-\frac{1}{6}\right)^2+
\frac{1}{180}\right]\left(\partial_i\partial_j-\delta_{ij}\nabla_x^2\right)
\int
d^3\tilde{x}\,\,\frac{R(\tilde{x})}{\left|\vec{x}-\vec{\tilde{x}}\right|^3}\,\,\,
,
\end{eqnarray}
therefore $\left\langle T_{\mu\nu}\right\rangle$ can be written in
terms of derivatives of $\left\langle \phi^2\right\rangle$. For a
spherically symmetric situation, using the series expansion Eq.
(\ref{eq;polenserie}) we find
\begin{eqnarray} \label{eq;Tmunuserie}
\left\langle T^t_{\phantom{t}t}\right\rangle &=& \frac{G}{2\pi^2}\left[\left(\xi-\frac{1}{6}\right)^2-
\frac{1}{90}\right]\sum_{n=1}^{\infty}n(2n+1)\frac{M_n}{r^{2n+3}}\nonumber\\
\left\langle T^r_{\phantom{r}r}\right\rangle &=& -\frac{G}{2\pi^2}\left[\left(\xi-\frac{1}{6}\right)^2+
\frac{1}{180}\right]\sum_{n=1}^{\infty}(2n+1)\frac{M_n}{r^{2n+3}}\nonumber\\
\left\langle T^{\theta}_{\phantom{\theta}\theta}\right\rangle &=&
\frac{G}{4\pi^2}\left[\left(\xi-\frac{1}{6}\right)^2+
\frac{1}{180}\right]\sum_{n=1}^{\infty}(2n+1)^2\frac{M_n}{r^{2n+3}}\nonumber\\\,\,
.
\end{eqnarray}
Note that the nonlocal contribution to $\left\langle
T^{\mu}_{\phantom{\mu}\nu}\right\rangle$ does not vanish for any
value of $\xi$.

The components of $\left\langle
T^{\mu}_{\phantom{\mu}\nu}\right\rangle$ fall off as $MG/r^5$ for
$r\gg R_0$. On the other hand, in the local approximation for the
Boulware vacuum (see for instance \cite{his},\cite{visser1}), they
fall off as $M^2G^2/r^6$. So the nonlocal approximation dominates
for $\left\langle T_{\mu\nu}\right\rangle$ as well as for
$\left\langle \phi^2\right\rangle$, by a factor of order $r/MG$.

Inside the star, $\left\langle T_{\mu\nu}\right\rangle$ can be
found by taking derivatives of the finite expression for
$\left\langle \phi^2\right\rangle$ (Eq.(\ref{eq;fourier})) in the
same fashion. As an example, the complete $\left\langle
T^{\mu}_{\phantom{\mu}\nu}\right\rangle$ for a star of constant
density is, for $r<R_0$,
\begin{eqnarray} \label{eq;Tmunuinterno}
\left\langle T^t_{\phantom{t}t}\right\rangle &=& -\frac{G\rho_0}{\pi}\left[\left(\xi-\frac{1}{6}\right)^2-
\frac{1}{90}\right]
\frac{3R_0^2-r^2}{(R_0^2-r^2)^2}\nonumber\\
\left\langle T^r_{\phantom{r}r}\right\rangle &=& -\frac{2G\rho_0}{\pi}\left[\left(\xi-\frac{1}{6}\right)^2+
\frac{1}{180}\right]\frac{1}{R_0^2-r^2}\nonumber\\
\left\langle T^{\theta}_{\phantom{\theta}\theta}\right\rangle &=&
-\frac{2G\rho_0}{\pi}\left[\left(\xi-\frac{1}{6}\right)^2+
\frac{1}{180}\right] \frac{R_0^2}{(R_0^2-r^2)^2} \,\, ,
\end{eqnarray}
while for $r>R_0$
\begin{eqnarray} \label{eq;Tmunuext}
\left\langle T^t_{\phantom{t}t}\right\rangle &=& \frac{G\rho_0}{\pi}\left[\left(\xi-\frac{1}{6}\right)^2-
\frac{1}{90}\right]
\frac{2R_0^3}{r(r^2-R_0^2)^2}\nonumber\\
\left\langle T^r_{\phantom{r}r}\right\rangle &=& -\frac{2G\rho_0}{\pi}\left[\left(\xi-\frac{1}{6}\right)^2+
\frac{1}{180}\right]\frac{R_0^2}{r^3(r^2-R_0^2)}
\nonumber\\
\left\langle T^{\theta}_{\phantom{\theta}\theta}\right\rangle &=&
\frac{2G\rho_0}{\pi}\left[\left(\xi-\frac{1}{6}\right)^2+
\frac{1}{180}\right] \frac{R_0^3(3r^2-R_0^2)}{2r^3(r^2-R_0^2)^2}
\,\, .
\end{eqnarray}
This result is quadratically divergent on the surface of the star.
To remove the logarithmic divergence in  $\left\langle
\phi^2\right\rangle$ it was enough to require $R(x)$ to be a
continuous function; as $\left\langle T_{\mu\nu}\right\rangle$ is
constructed with second derivatives of $\left\langle
\phi^2\right\rangle$, to ensure that no divergence arises on the
surface it is necessary to require the continuity of the second
derivatives of $R(x)$.

The local approximation for $\left\langle T_{\mu\nu}\right\rangle$
inside a star of constant density, presented in \cite{his}, gives
an expression with an additional factor of $MG/R_0$. Therefore the
nonlocal approximation is, for nonrelativistic stars, dominant
over the local one both in the internal as well as the external
region.

\section{Applications}
\subsection{Energy conditions}

One important question to ask is whether $\left\langle
T_{\mu\nu}\right\rangle$ calculated within semiclassical gravity
satisfies the energy conditions or not. It is well known that
quantum fields often do not satisfy the energy  conditions
\cite{eps}, at least not in the same way as classical fields do.
This fact has raised considerable discussion regarding its
implications for singularity theorems, black hole dynamics,
existence  of macroscopic traversable wormholes, creation of
closed time-like curves, etc \cite{morris}. In other words,
violation of the energy conditions imply that the semiclassical
Einstein equations, with  $\left\langle T_{\mu\nu}\right\rangle$
as a source, could in principle admit solutions qualitatively
different from classical solutions \cite{flana}.

In this section we will discuss the validity of two energy
conditions, the \textit{Null Energy Condition} (NEC) and the
\textit{Averaged Null Energy Condition} (ANEC) \cite{anec}, for
the massless scalar field $\left\langle T_{\mu\nu}\right\rangle$
found in last section. The discussion will be restricted to the
exterior region, since inside the star all energy conditions are
satisfied by the classical source. These conditions have been
studied in Ref.  \cite{visser1} using the local approximation for
the Boulware vacuum in the conformal case. The conclusion was that
neither of them holds in the exterior of a non-collapsed star.
However, since the nonlocal contribution to $\left\langle
T_{\mu\nu}\right\rangle$ dominates in the weak field limit,  it is
necessary to revise the calculation.

The NEC states that $T_{\mu\nu}K^{\mu}K^{\nu}\geq 0$ for every
null vector $K^{\mu}$. For a static spherically symmetric
situation this condition is equivalent to the following
inequalities
\begin{equation}
\left\langle T^{t}_{\phantom{t}t}\right\rangle \geq \left\langle
T^{r}_{\phantom{r}r}\right\rangle  \quad \left\langle
T^{t}_{\phantom{t}t}\right\rangle \geq \left\langle
T^{\theta}_{\phantom{\theta}\theta}\right\rangle\,\,\, ,
\end{equation}
to which we will refer as radial and tangential conditions
respectively. Using the series expansions (\ref{eq;Tmunuserie}) we
find that
\begin{equation}
\left\langle T^{t}_{\phantom{t}t}\right\rangle - \left\langle
T^{r}_{\phantom{r}r}\right\rangle =
\frac{G}{2\pi^2}\sum_{n=1}^{\infty}\frac{M_n}{r^{2n+3}}\left(2n+1\right)\times
\left[-\frac{1}{90}\left(n-\frac{1}{2}\right)+\left(\xi-\frac{1}{6}\right)^2\left(n+1\right)\right]
\end{equation}

\begin{equation}
\left\langle T^{t}_{\phantom{t}t}\right\rangle - \left\langle
T^{\theta}_{\phantom{\theta}\theta}\right\rangle=-\frac{G}{2\pi^2}\sum_{n=1}^{\infty}\frac{M_n}{r^{2n+3}}\left(2n+1\right)
\times
\left[\frac{1}{360}\left(6n+1\right)+\frac{1}{360}\left(\xi-\frac{1}{6}\right)^2\right]\,\,
.
\end{equation}
It is easily seen that the tangential condition is violated for
any $\xi$, which is enough to ensure NEC violation. The radial
condition is also violated in the conformal case, but it holds in
the minimal coupling case. Therefore the possibility remains that
in this case the average of the NEC condition over a null geodesic
may be positive. We will show now that this is not the case: the
ANEC is also violated for arbitrary $\xi$.

The ANEC \cite{anec} states that
$\int_{-\infty}^{\infty}dp\,\,T_{\mu\nu}K^{\mu}K^{\nu}\geq 0$ for
any null geodesic of tangent vector $K^{\mu}$ and affine parameter
$p$. To prove that the ANEC is also violated, we consider a null
geodesic external to the star with impact parameter $b > R_0$,
defined by the following functions $x^{\mu}(p)$
\begin{eqnarray} \label{eq;parametrizacion}
t(p) & = & p \,\,\, \quad
r(p)  =  \sqrt{p^2+b^2} \nonumber\\
\theta (p) & = & \pi/2 \,\,\, \quad \varphi (p)  =  \arctan
(p/b)\,\, .
\end{eqnarray}
The geodesic is taken to be a straight line, to keep the analysis
consistently at first order. The integration of $\left\langle
T_{\mu\nu}\right\rangle K^{\mu}K^{\nu}$ over the whole trajectory
(with $K^{\mu}=\frac{\mathrm{d}x^{\mu}}{\mathrm{dp}}$) can be
performed using the series expansion for $\left\langle
T_{\mu\nu}\right\rangle $ in Eq.(\ref{eq;Tmunuserie}). The result
is
\begin{eqnarray}
\lefteqn{\int_{-\infty}^{\infty}dp\,\,\left\langle T_{\mu\nu}\right\rangle K^{\mu}K^{\nu}=-
\frac{G}{2880\, b^2}\sum_{n=1}^{\infty}\frac{M_n}{(2b)^{2n}}\times}\nonumber\\
&&\left(2n+1\right)\frac{2\left(4n(n+1)-3\right)\Gamma(2n+2)+\Gamma(2n+4)}{\Gamma(n+3/2)\,\Gamma(n+5/2)}\,\,
.
\end{eqnarray}
This quantity is clearly negative and independent of $\xi$. We conclude that the ANEC is violated for a
massless scalar field in the exterior of a nonrelativistic star independently both of the coupling of the field
and of the internal structure of the star.

\subsection{Quantum corrections to the metric}

In this section we shall solve the linearized semiclassical
Einstein equations, which take into account the backreaction of
the quantum field over the spacetime metric. The results will be
found to agree with those obtained by \cite{newton} but disagree
with those found relying on the local approximation for
$\left\langle T_{\mu\nu}\right\rangle$ \cite{his}.

Disregarding both cosmological constant and higher-order terms,
the semiclassical Einstein equations read
\begin{equation} \label{eq;einstein}
R_{\mu\nu} -\frac{1}{2}R g_{\mu\nu} =  -8\pi G
\Big(T_{\mu\nu}^{(\mathrm{cl})}+\left\langle
T_{\mu\nu}\right\rangle\Big) \,\, .
\end{equation}
We take the source term to be separated in a classical part, the
ordinary density of the star, and a quantum part, which is the
$\left\langle T_{\mu\nu}\right\rangle$ we have calculated in the
previous section. The gravitational field is similarly separated
in a classical term (the background spacetime) and a correction
that is produced by the quantum $\left\langle
T_{\mu\nu}\right\rangle$
\begin{equation}
g_{\mu\nu}=\eta_{\mu\nu}+h_{\mu\nu}^{(\mathrm{cl})}+h_{\mu\nu}^{(\mathrm{q})}
\,\, .
\end{equation}

We express the correction to the gravitational field in isotropic
form. In Cartesian coordinates
\begin{equation}
h_{00}^{(\mathrm{q})}=f(r)\,,\quad\quad
h_{0i}^{(\mathrm{q})}=0\,,\quad\quad
h_{ij}^{(\mathrm{q})}=g(r)\delta_{ij} \,\, .
\end{equation}
It is a simple matter to solve the equations (\ref{eq;einstein}),
linearized in $h_{\mu\nu}$, using the $\left\langle
T_{\mu\nu}\right\rangle$ given by Eq. (\ref{eq;Tmununoloc}) as a
source. The result is
\begin{equation} \label{eq;fesferica}
f(r)=-\frac{G}{4\pi^2}\left[\left(\xi-\frac{1}{6}\right)^2+
\frac{1}{45}\right]\int d^3\tilde{x}\,\,\frac{R(\tilde{x})}{\left|\vec{x}-\vec{\tilde{x}}\right|^3}
\end{equation}
\begin{equation} \label{eq;gesferica}
g(r)=-\frac{\left[\left(\xi-\frac{1}{6}\right)^2-
\frac{1}{90}\right]}{\left[\left(\xi-\frac{1}{6}\right)^2+
\frac{1}{45}\right]}f(r) \,\, .\end{equation} Note that the
correction $h_{\mu\nu}^{(\mathrm{q})}$ is proportional to
$\left\langle \phi^2\right\rangle$ (except in the conformal case
$\xi=1/6$, in which $\left\langle \phi^2\right\rangle$ vanishes
but $h_{\mu\nu}^{(\mathrm{q})}$ does not). Eq.
(\ref{eq;fesferica}) is correct only outside the star, where
$\left\langle \phi^2\right\rangle$ is given by Eq.
(\ref{eq;GFnoloc}). Inside the star, $h_{\mu\nu}^{(\mathrm{q})}$
has the same proportionality with the finite $\left\langle
\phi^2\right\rangle$ given in Eq. (\ref{eq;fourier}).

In the exterior region we can replace Eq.(\ref{eq;fesferica}) by
its multipolar expansion
\begin{equation} \label{eq;fenserie}
f(r)=-\frac{2G^2}{\pi}\left[\left(\xi-\frac{1}{6}\right)^2+
\frac{1}{45}\right]\sum_{n=1}^{\infty}\frac{M_n}{r^{2n+1}} \,\, .
\end{equation}
Here we see an explicit dependence of the external metric with the
internal structure of the star, which can be thought as a "quantum
violation" of Birkoff's theorem. This is possible because the
space surrounding the star, although empty at the classical level,
contains a quantum vacuum energy given by $\left\langle
T_{\mu\nu}\right\rangle$.

As an example, the quantum correction to the Schwarzschild metric
external to a star of constant density is predicted by our
nonlocal approximation to be
\begin{equation} \label{eq;frhocte}
f(r)=-\frac{2}{\pi}\left[\left(\xi-\frac{1}{6}\right)^2+
\frac{1}{45}\right]\frac{MG^2}{r^3}\left[1+\frac{3}{5}\frac{R_0^2}{r^2}+\cdots
\right] \,\, ,
\end{equation}
with $g(r)$ given by Eq. (\ref{eq;gesferica}).

The leading term in Eqs. (\ref{eq;fenserie}) and
(\ref{eq;frhocte}) for $r\gg R_0$ is the quantum correction to the
Newtonian potential of a point mass, since it contains no
information about the internal structure. Therefore the complete
Newtonian potential of a point mass when the effect of a quantum
massless scalar field is taken into account reads
\begin{equation} \label{eq;potnewtoniano}
\Phi(r)=-\frac{MG}{r}-\frac{1}{\pi}\left[\left(\xi-\frac{1}{6}\right)^2+
\frac{1}{45}\right]\frac{MG^2}{r^3} \,\, .
\end{equation}
This result agrees with previous ones \cite{newton}. It also has
the same form that the long-distance quantum corrections to $\Phi$
due to gravitons, calculated in the low energy limit of quantum
gravity \cite{dono}. By contrast, the correction to the
Schwarszchild metric found in \cite{his} using the local
approximation for $\left\langle T_{\mu\nu}\right\rangle$ is
\begin{equation}\label{eq;potnewtonhiscock}
g_{00}(r\rightarrow\infty)=1-\frac{2MG}{r}-\frac{M^2G^2}{60\pi
r^4}+\cdots \,\,\, ,
\end{equation}
which does not agree with previous results.

It is worth noting that our result satisfies consistently the
principle of superposition; if the gravitational field of the star
is calculated as a sum of infinitesimal contributions of the form
given in Eq. (\ref{eq;potnewtoniano}), the result takes us back to
Eq. (\ref{eq;fesferica}). In fact, Eq. (\ref{eq;fesferica}) is
correct even if there is no spherical symmetry, as can be seen by
noting that the $00$ component of the linearized Einstein
equations reads (in the Lorentz gauge)
\begin{equation}
\square h_{00}^{(\mathrm{q})}=-16\pi G \left(\left\langle
T_{00}\right\rangle- \frac{1}{2} \left\langle
T^{\lambda}_{\phantom{\lambda}\lambda}\right\rangle\right) \,\, .
\end{equation}
Using the expressions Eq. (\ref{Tmunuderivado}) for $\left\langle
T_{\mu\nu}\right\rangle$, it follows immediately that
$h_{00}^{(\mathrm{q})}(x)$ is given by Eq. (\ref{eq;fesferica})
(if the point $x$ is located outside the sources, i.e., if
$R(x)=0$).

As an application in a non-spherically symmetric situation, we
outline here the calculation of the long-distance quantum
correction to the Kerr metric by considering a rotating source.
The classical $T_{\mu\nu}$ has now off-diagonal terms related to
the angular velocity of the star, $T_{0i}(x)=\rho \omega r\cos
\theta (-\sin \varphi \hat{x}+\cos \varphi \hat{y})_i$. This
produces a nonzero $H_{0i}^{(2)}$ which in turn causes
$\left\langle T_{\mu\nu}\right\rangle$ to have, outside the star,
the following components
\begin{equation}
\left\langle T_{0i}\right\rangle=\frac{G\omega}{240
\pi^2}\nabla^2\,\int \mathrm{d^3}x'\, \frac{\rho(x')r'\cos \theta'
(-\sin \varphi' \hat{x}+\cos \varphi'
\hat{y})_i}{\left|\vec{x}-\vec{x'}\right|^3} \,\, ,
\end{equation}
in addition to those included in Eq. (\ref{Tmunuderivado}), which
are not modified by the rotation. Inserting $\left\langle
T_{0i}\right\rangle$  in the semiclassical Einstein equations, we
find that the long-distance correction to the off-diagonal terms
of the Kerr metric are of the form
\begin{equation}
h_{0i}=\frac{G^2}{10\pi r^5}(\textbf{J}\times
\textbf{r})_i+O\left( \frac{G^2JR_0^2}{r^6}\right) \,\, ,
\end{equation}
where \textbf{J} is the total angular momentum of the source. Note
that the information about internal structure is again contained
in the sub-dominant term, which always falls off two powers of $r$
faster than the dominant one. Once again, similar results hold for
the graviton quantum corrections \cite{dono2}.

\section{Conclusions}

In this paper we computed $\langle \phi^2\rangle$ and $\langle
T_{\mu\nu}(x)\rangle$ for a massless quantum field with arbitrary
coupling to the curvature in the geometry of spherically symmetric
Newtonian stars. We obtained  expressions for both quantities as
multipolar expansions, valid in the weak field limit, that show
explicitly the nonlocal dependence of the quantum effects. As a
byproduct, we have demonstrated that previous local approximations
to the vacuum polarization in spherically symmetric spacetimes do
not apply to Newtonian stars. Indeed, the nonlocal part is the
leading contribution that overwhelms the local one both inside and
outside the surface of the star. The results are divergent on the
surface if the star model is not taken as sufficiently smooth.

We have shown that outside the star $\langle T_{\mu\nu}(x)\rangle$
violates the NEC and ANEC energy conditions. A local approximation
for $\langle T_{\mu\nu}(x)\rangle$ was used in previous proofs of
these results, which stand now on firmer grounds. Our results hold
for fields with arbitrary coupling, while previously only
conformal fields had been considered. We also computed the quantum
corrections to the metric, and found a "quantum violation" to
Birkoff's theorem: the external metric depends not only on the
mass but on all the multipolar moments of the distribution within
the star. From the same principles it is also easily shown that,
due to quantum corrections, the gravitational field inside a
spherical shell is slightly different from zero.

We have pointed out that the nonlocal quantum correction to the
metric outside the star is a consequence of the $1/r^3$
modification to the Newtonian potential and the superposition
principle. This fact can be used to argue that the nonlocal
modifications to the metric will  be present in any quantum theory
of gravity, since gravitons also induce $1/r^3$-corrections to the
Newtonian potential. These corrections are, of course, extremely
small.

\section{Acknowledgments}
F.D.M. was supported by Universidad de Buenos Aires, CONICET, and
Agencia Nacional de Promoci\'on Cient\'\i fica y Tecnol\'ogica,
Argentina. E.A. would like to thank the warm hospitality of the
University of Buenos Aires.

\end{document}